\pdfoutput=1
\documentclass[a4paper]{llncs}[2018/03/10]
\usepackage{mathptmx}
\usepackage{helvet}
\usepackage{courier}
\usepackage{type1cm}
\usepackage{makeidx}
\usepackage[rgb,dvipsnames,x11names]{xcolor}
\usepackage{graphicx}
\usepackage{hyperref}
\usepackage[all]{hypcap}
\usepackage{bookmark}
\usepackage{amsfonts}
\usepackage[cmex10]{amsmath}
\usepackage{amssymb}

\DeclareGraphicsExtensions{.mps}
\hypersetup{%
	pdfdisplaydoctitle=true,
	pdftitle={On the Statistics of Urban Street Networks},
	pdfauthor={J\'er\^ome~Benoit (ORCid: 0000-0003-1226-6757) and Saif~Eddin~Jabari},
	pdfsubject={ComplexNetwork 2018
		- The 7th International Conference on Complex Networks and Their Applications
		- Cambridge 11-13 December (United Kingdom)%
		},
	pdfkeywords={
		extended abstract
		},
	pdfcreator={\LaTeXe{} and its friends},
	pdfpagelayout=SinglePage,
	pdfpagemode=UseOutlines,
	pdfstartpage=1,
	pdfhighlight=/O,
	pdfview=FitH,
	pdfstartview=FitH,
	colorlinks=true,
	allcolors=RedOrange,
	citecolor=RoyalBlue3,
	urlcolor=RoyalBlue3,
	linkcolor=RoyalBlue3,
	bookmarksnumbered=true,
	bookmarksopen=true,
	bookmarksopenlevel=3,
	}

\DeclareMathOperator{\sfhzeta}{\zeta}
\DeclareMathOperator{\sfWitten}{\mathcal{W}}

\begin{document}

\title{On the Statistics of Urban Street Networks}
\titlerunning{On the Statistics of Urban Street Networks}
\author{%
	J\'er\^ome Benoit
	\and
	Saif Eddin Jabari
	}
\authorrunning{J\'er\^ome Benoit and Saif Eddin Jabari}
\tocauthor{J\'er\^ome Benoit and Saif Eddin Jabari}
\institute{%
	New~York~University Abu~Dhabi,
	Abu~Dhabi,
	United~Arab~Emirates\\
	\email{jerome.benoit@nyu.edu}\\
	}
\maketitle

\section{Introduction}

We seek to understand the statistics of urban street networks.
Such an understanding will improve urban policies in general and urban transportation in particular.
In our work here we investigate urban street networks as a whole
within the frameworks of information physics \cite{KHKnuth2011}
and statistical physics \cite{ETJaynes1957I}.

Although
the number of times that a \emph{natural road} crosses an other one
has been widely observed to follow a
discrete Pareto probability distribution \cite{AClausetCRShaliziMEJNewman2009}
among self-organized cities \cite{CAlexanderACINAT1965,CrucittiCMSNUS2006,BJiangTSUSNPDC2014},
very few efforts have focused
on deriving
the statistics of urban street networks
from fundamental principles.
Here a natural road (or road) denotes an accepted substitute for a ``named'' street \cite{BJiangTSUSNPDC2014}.

Our approach explicitly emphasizes
the road-junction hierarchy of the initial urban street network
rather than implicitly splitting it accordingly in two dual but distinct networks.
Most of the investigations indeed seek
to cast the initial urban street network into a road-road
network \cite{BJiangTSUSNPDC2014}
and to describe its valence probability distribution.

This holistic viewpoint adopted by the urban community \cite{CAlexanderACINAT1965,RHAtkin1974}
also
fits with the mindset of information physics \cite{KHKnuth2011},
which is built upon partial order relations \cite{BADaveyHAPriestleyILO,KHKnuth2011}.
Here
the partial order relation derives from the road-junction incidence relation.
Applying then information physics
enables us to envisage urban street networks as evolving social systems
subject to an entropic equilibrium
similar to the Paretian one effectively observed among cities of a same country
\cite{YDover2004}.

\section{Method}

The relation that ties natural roads and junctions is bijectively reduced
into an algebraic structure known as Galois lattice \cite{BADaveyHAPriestleyILO}.
Then,
by imposing natural consistency constraints,
information physics \cite{KHKnuth2011}
enables us not only to evaluate
urban street networks
but also to assess
a probability distribution and, ultimately,
an information measure.
It appears that
urban street networks effortlessly reduce to
Galois lattices with two nontrivial layers:
the natural roads form the lower layer
and the junctions form the upper one,
while the partial ordering relation is ``passing through.''
This causes urban street networks to become a \textit{toy model} for the emerging paradigm.

The passage from Galoisean hierarchy to Paretian coherence
is then achieved by invoking Jaynes's Maximum Entropy principle \cite{ETJaynes1957I}
with the first logarithmic moment as the sole characterizing constraint
and our complete ignorance as initial knowledge.
The corresponding most plausible probability distribution
expresses
for each natural road or junction,
indiscriminately,
the likeliness of possessing a given number of equally likely states;
it is a discrete Pareto probability distribution
whose entropy is the imposed first logarithmic moment,
as desired.
This probability distribution must ultimately be decomposed
with respect to the structure of the Galois lattice
and the algebraic rules imposed by information physics theory.
In other words,
a physical meaning remains to be given to the evaluation.

Hypothesizing a crude asymptotic binomial paired-agent
model with the spirit of the city model \cite{YDover2004}
allows us finally to predict the statistics of urban street networks.
Here each natural road or junction is envisioned as
an intranetwork whose very survival relies
on the ability of each of its agent
to preserve a crucial number of intraconnections.
Thereby each urban street network becomes characterized
by two generalized binomial combination numbers which asymptotically rise up
as two characterizing exponents beside the Paretian exponent $\lambda$:
the \emph{numbers of vital connections} for natural roads
and for junctions,
respectively $\upsilon_{r}$ and $\upsilon_{j}$.

\begin{figure}[bh]
	\begin{center}
		\includegraphics[width=0.85\linewidth]{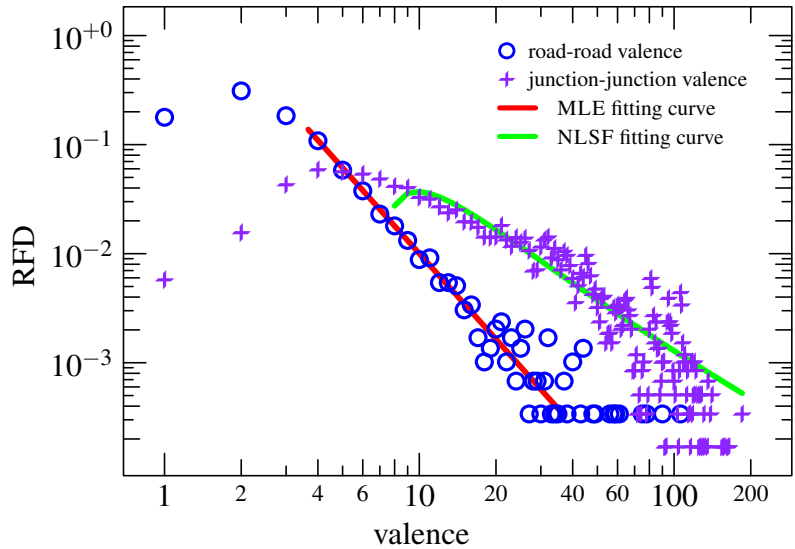}
	\end{center}
	\caption{\label{OSUSN/fig/USN/London}%
		Relative Frequency Distributions (\textsc{RFD}) for the urban street network of London:
		circles represent relative frequencies for the valences of the road-road topological network;
		crosses represent relative frequencies for the valences of the junction-junction topological network.
		The red fitted curve
		for the natural road statistics
		describes the Maximum Likelihood Estimate (\textsc{MLE})
		for the discrete Pareto probability distribution \eqref{OSUSN/eq/USN/PDF/NaturalRoads}
		estimated according to the state of the art \cite{AClausetCRShaliziMEJNewman2009}
		(%
			$\underline{n}_{r} = 4$,
			$2\lambda\upsilon_{r} = 2.610(0.065)$,
			$n=5000$ samples,
			$p\text{-value} = 0.929$%
			).
		The green fitted curve for the junction statistics
		shows the best Nonlinear Least-Squares Fitting (\textsc{NLSF})
		for the nonstandard discrete probability distribution \eqref{OSUSN/eq/USN/PDF/Junctions}
		with $\underline{n}_r$ and $2\lambda\upsilon_{r}$ fixed to
		their respective \textsc{MLE} value ($2\lambda\upsilon_{j} = -1.3$);
		since
		fast evaluation of the normalizing function $\sfWitten$ has yet to be found,
		no \textsc{MLE} approach can be used for now.
		Having a number of vital connections $\upsilon_{j}$ negative
		means that the associated generalized binomial combination number is smaller than one,
		{i.e.},
		that the number of agent intraconnections for junctions
		is relatively much smaller than the one for natural roads.%
		}
\vspace{-0.4cm}
\end{figure}

\section{Results and Discussion}

Our approach
recovers
the discrete Pareto probability distribution widely observed
for natural roads evolving in self-organized cities,
and foresees
a nonstandard bell-shaped distribution with a Paretian tail for their joining junctions.
The probability for a natural road to cross ${n}_{r}$ natural roads is
\begin{subequations}\label{OSUSN/eq/USN/PDF}
\begin{equation}\label{OSUSN/eq/USN/PDF/NaturalRoads}
	\Pr\left({n}_{r}\right) =
		\frac{{n}_{r}^{-2\lambda\upsilon_{r}}}{\sfhzeta\left(2\lambda\upsilon_{r};\underline{n}_{r}\right)}
\end{equation}
where
\begin{math}
	\sfhzeta\left(\alpha;\underline{n}\right) =
		\sum_{{n}={n}_r}^{\infty} {n}^{-\alpha}
\end{math}
is the generalized zeta function,
and
the probability for a junction to see ${n}_{j}$ junctions
through its joining natural roads reads
\begin{equation}\label{OSUSN/eq/USN/PDF/Junctions}
	\Pr\left({n}_{j}\right) =
		\frac{%
			\sum_{{n}=\underline{n}_{r}}^{{n}_{j}-\underline{n}_{r}}
				\left[{n}\left({n}_{j}-{n}\right)\right]^{-2\lambda\upsilon_{r}}
				\,%
				{n}_{j}^{-2\lambda\upsilon_{j}}
			}{%
			\sfWitten\left(2\lambda\upsilon_{r},2\lambda\upsilon_{r},2\lambda\upsilon_{j};\underline{n}_{r}\right)%
			}
\end{equation}
where
\begin{math}
	\sfWitten\left(\alpha,\beta,\gamma;\underline{n}\right) =
		\sum_{{m},{n}\geqslant\underline{n}}
			{m}^{-\alpha} {n}^{-\beta} \left(m+n\right)^{-\gamma}
\end{math}
is the two-dimensional
generalized
Mordell-Tornheim-Witten zeta function;
\end{subequations}
the number of junctions per natural road ${n}_{r}$ is assumed
to span from some minimal value $\underline{n}_{r}$
for practical reasons \cite{AClausetCRShaliziMEJNewman2009}.

Figure~\ref{OSUSN/fig/USN/London} exhibits
the urban street network of London as a case study.
The probability distribution for natural roads
$\Pr\left({n}_{r}\right)$ \eqref{OSUSN/eq/USN/PDF/NaturalRoads} is highly plausible,
as expected for any recognized self-organized city \cite{CAlexanderACINAT1965,BJiangTSUSNPDC2014}.
The validation of the probability distribution for junctions $\Pr\left({n}_{j}\right)$ \eqref{OSUSN/eq/USN/PDF/Junctions}
appears more delicate for the time being.
Meanwhile a crude data analysis is not conclusive enough.
Interestingly,
this case study reveals that
the number of intraconnections for junctions
might be relatively much smaller than the one for natural roads
in self-organized cities.

Thus the statistical model for urban street networks \eqref{OSUSN/eq/USN/PDF}
appears fine enough
to study urban macro behaviours
with the exponents $\lambda$, $\upsilon_{r}$, and $\upsilon_{j}$ as complexity parameters.
Future work includes
\newcounter{counterFWEnum}\setcounter{counterFWEnum}{1}%
(\roman{counterFWEnum})\stepcounter{counterFWEnum}~%
finding patterns {via} the ratio ${\upsilon_{r}}/{\upsilon_{j}}$
among
self-organized cities,
(\roman{counterFWEnum})\stepcounter{counterFWEnum}~%
extending the model to
designed cities,
(\roman{counterFWEnum})\stepcounter{counterFWEnum}~%
applying the
paradigm to more intricate systems,
and
(\roman{counterFWEnum})\stepcounter{counterFWEnum}~%
full
investigation of the resulting Paretian statistical physics.

\bibliographystyle{splncs03}

\end{document}